\documentclass[conference]{IEEEtran}

\def\BibTeX{{\rm B\kern-.05em{\sc i\kern-.025em b}\kern-.08em
		T\kern-.1667em\lower.7ex\hbox{E}\kern-.125emX}}

\usepackage{xspace}
\usepackage{syntax}
\usepackage{listings}
\usepackage{xcolor}
\usepackage{multicol}
\usepackage{multirow}
\usepackage{array}
\usepackage{pifont}
\usepackage{amsmath}
\usepackage{algpseudocode}
\usepackage{algorithm}
\usepackage{tikz}
\newcommand*\circled[1]{\tikz[baseline=(char.base)]{
                \node[shape=circle,fill,inner sep=0.5pt] (char) {\bfseries\footnotesize\textcolor{white}{#1}};}}
 
\newcommand*\halfcirc[1][1ex]{\begin{tikzpicture}[baseline=(char.base), inner sep=0.5pt]
  \draw[fill] (0,0)-- (90:#1) arc (90:270:#1) -- cycle ;
  \draw (0,0) circle (#1);
  \end{tikzpicture}}
\newcommand*\fullcirc[1][1ex]{\tikz[baseline=(char.base), inner sep=0.5pt]\fill (0,0) circle (#1);} 
\newcommand{\cmark}{\ding{51}}\newcommand{\xmark}{\ding{55}}\newcommand{\smark}{\ding{73}}

\usepackage{enumitem}

\definecolor{codegreen}{rgb}{0,0.6,0}
\definecolor{codegray}{rgb}{0.5,0.5,0.5}
\definecolor{codepurple}{rgb}{0.58,0,0.82}
\definecolor{backcolour}{rgb}{0.95,0.95,0.92}
\lstdefinestyle{TAPstyle}{
	backgroundcolor=\color{backcolour},   
    commentstyle=\color{codegreen},
    keywordstyle=\color{magenta},
    numberstyle=\tiny\color{codegray},
    stringstyle=\color{codepurple},
    basicstyle=\ttfamily\footnotesize,
    breakatwhitespace=false,         
    breaklines=true,                 
    captionpos=b,                    
    keepspaces=false,                 
numbersep=0pt,                  
    showspaces=false,                
    showstringspaces=false,
    showtabs=false,                  
    tabsize=2,
    xleftmargin=5pt,
    xrightmargin=5pt,
morekeywords={}
}
\usepackage[numbers,sort&compress]{natbib}
\usepackage[hidelinks=true,
plainpages=false,
breaklinks=true
]{hyperref}
\hypersetup{pdfsubject={},
	pdfauthor={},
	pdftitle={},
	pdfkeywords={}
} 

\usepackage[T1,OT1]{fontenc}

\newcommand{\ie}{\textit{i.e.}\xspace}
\newcommand{\eg}{\textit{e.g.}\xspace}

\newcommand{\code}[1]{{\texttt{#1}}\xspace}
\newcommand{\FIGREF}{Figure\xspace}

\newcommand{\PKCSone}{PKCS\#1-v1.5\xspace}
\newcommand{\pkcs}{PKCS\#1\xspace}
\newcommand{\afl}{AFL\xspace}
\newcommand{\aflplusplus}{AFL++\xspace}
\newcommand{\honggfuzz}{HonggFuzz\xspace}
\newcommand{\libfuzzer}{libFuzzer\xspace}
\newcommand{\peach}{Peach\xspace}
\newcommand{\morpheus}{Morpheus\xspace}
\newcommand{\isla}{ISLA\xspace}
\newcommand{\darwin}{Darwin\xspace}
\newcommand{\aflsmart}{AFLSmart\xspace}
\newcommand{\nautilus}{Nautilus\xspace}
\newcommand{\sgfuzz}{SGFuzz\xspace}

\newcommand{\editdist}{\ensuremath{\mathcal{D}}\xspace}
\newcommand{\lcs}{\ensuremath{\mathcal{L}}\xspace}
\newcommand{\lcsfunc}{\texttt{{lcs}\xspace}}

\renewcommand\appendix{\par
	\setcounter{section}{0}
	\setcounter{subsection}{0}
	\setcounter{figure}{0}
	\setcounter{table}{0}
	\setcounter{lstlisting}{0}
        \renewcommand\thesection{\Alph{section}}
	\renewcommand\theHsection{\Alph{section}}
	\renewcommand\thefigure{\Alph{section}\arabic{figure}}
	\renewcommand\theHfigure{\Alph{section}\arabic{figure}}
	\renewcommand\thetable{\Alph{section}\arabic{table}}
	\renewcommand\theHtable{\Alph{section}\arabic{table}}
	\renewcommand\thelstlisting{\Alph{section}\arabic{lstlisting}}
	\renewcommand\theHlstlisting{\Alph{section}\arabic{lstlisting}}
}

\begin{document}

\title{FuzzEval: Assessing Fuzzers on Generating Context-Sensitive Inputs}

\author{
	\IEEEauthorblockN{S Mahmudul Hasan,
		Polina Kozyreva, and Endadul Hoque}
	\IEEEauthorblockA{\textit{Syracuse University, NY USA}}
}

\maketitle

\begin{abstract}
Cryptographic protocols form the backbone of modern security systems, 
yet vulnerabilities persist within their implementations. 
Traditional testing techniques, including fuzzing, have struggled to 
effectively identify vulnerabilities in cryptographic libraries due to their 
reliance on context-sensitive inputs. 
This paper presents a comprehensive evaluation of eleven state-of-the-art 
fuzzers' ability to generate context-sensitive inputs for testing 
a cryptographic standard, \PKCSone, across thirteen implementations. 
Our study reveals nuanced performance differences among the fuzzers in terms of
the validity and diversity of the produced inputs.
This investigation underscores the limitations of existing fuzzers
in handling context-sensitive inputs. 
These findings are expected to drive further research and development in
this area. 

\end{abstract}

\begin{IEEEkeywords}
Fuzzing, Cryptographic protocols, \PKCSone, and context-sensitive inputs
\end{IEEEkeywords}

\section{Introduction}
\label{sec:intro}

Cryptographic protocols play an unparalleled role in security. Despite their importance, many libraries that implement these protocols still contain numerous vulnerabilities, as disclosed by specialized testing tools
\cite{chau2019analyzing,yahyazadeh2021morpheus,chau2017symcerts,brubaker2014frankencerts}. 
While fuzzing, a widely employed security testing technique, has successfully unearthed numerous flaws, even robust general-purpose fuzzers like AFL and its variants \cite{fuzzing2021,fuzzingsurvey,yahyazadeh2021morpheus} 
struggle to detect vulnerabilities in cryptographic libraries. This raises a crucial question: 
What specific limitations hinder these fuzzers from effectively identifying security 
vulnerabilities in cryptographic libraries?
This inquiry serves as the central focus of our paper.

At first glance, one might argue that general-purpose fuzzers like \afl excel in 
finding memory-safety bugs but lack the capability to effectively monitor 
their fuzz targets, such as cryptographic libraries, for the same types of bugs 
that custom tools specialize in. 
While this assumption appears valid, it has been observed that many custom-made tools 
have also successfully uncovered memory-safety bugs overlooked by \afl-like fuzzers \cite{AFLvariants}, 
despite their reputation for detecting such vulnerabilities 
\cite{yahyazadeh2021morpheus,brubaker2014frankencerts}. 
This observation raises a deeper question: How do custom-made tools, exemplified by \morpheus \cite{yahyazadeh2021morpheus}, 
exhibit these behaviors within the target, while AFL does not?

The key to manifesting such bugs lies in the necessity for test inputs 
that not only carry specific values but also pass stringent input validation, 
a common requirement for cryptographic protocols \cite{rfc2313,TLS1.2,sshrfc}. 
These input validations entail intricate sanity checks embedded within the code 
to ensure that only correctly formatted inputs are processed by the implementation. 
An input is considered correctly formatted, aka \textit{\textbf{valid}},
when it satisfies the required semantic properties, 
as dictated by the protocol standards. Semantic properties of cryptographic protocol inputs 
are often \textit{\textbf{context-sensitive}}, as they express intricate relationships 
between different parts, also known as \textit{fields}, of the input.

In essence, uncovering such bugs requires generating valid context-sensitive test inputs 
capable of navigating stringent input validation and reaching the locations of the bugs. 
While custom tools like \morpheus excel at generating these test inputs, 
general-purpose fuzzers like \afl struggle to generate inputs essential for testing 
beyond initial input validation, consequently missing such implementation flaws \cite{yahyazadeh2021morpheus}. 
This suggests that a fuzzer capable of generating semantically valid inputs 
has a higher likelihood of uncovering such bugs.

This realization prompted us to undertake a broader investigation. 
While a variety of general-purpose fuzzers exist, differing in their approach 
from black-box to grey-box and utilizing techniques such as input grammars or seed inputs, 
few explicitly address the need for context-sensitive inputs \cite{steinhofel2022input,fuzzingsurvey}. 
To assess the suitability of these state-of-the-art general-purpose fuzzers 
for testing cryptographic libraries, we focus on evaluating their capability 
to generate semantically valid inputs, which forms the crux of our study.

Prior work has established guidelines for evaluating fuzzers 
\cite{klees2018evaluating,fuzzingsurvey,bundt2021evaluating, gorz2023systematic}, 
with a primary focus on assessing their capability to discover known bugs and 
their achieved performance in terms of coverage-based metrics, such as branch coverage. 
However, these studies overlook assessing the suitability of fuzzers for testing implementations 
like cryptographic libraries, which heavily rely on context-sensitive inputs.

In this paper, we conducted a large-scale study to evaluate 
eleven state-of-the-art fuzzers' ability to generate semantically valid context-sensitive inputs 
for fuzzing a cryptographic library. These fuzzers, 
including \afl \cite{afl}, \aflplusplus \cite{AFLplusplus-Woot20}, 
\libfuzzer \cite{libfuzzer}, \nautilus \cite{aschermann2019nautilus}, and others, 
employ various mechanisms such as grey-box and black-box approaches, as well as input grammars. 
Our target for fuzzing was the \PKCSone cryptographic standard for RSA signature schemes \cite{rfc2313}. 
We chose \PKCSone for its widespread use in secure communication protocols (\eg, SSL/TLS, IPSec), 
software signing, and X.509 certificates. Given its importance and the availability of numerous implementations suitable for fuzzing, we evaluated the implementations of \PKCSone signature verification from 13 popular cryptographic libraries, 
such as OpenSSL.

To ensure comparability and reproducibility in fuzzer evaluation, 
we developed a Linux-container-based testing platform for each $\langle$fuzzer, test-subject$\rangle$ pair.
The platform includes a central controller that orchestrates the fuzzing campaign and 
collects inputs generated by the fuzzers through a locally hosted TCP server. 
The server employs a \PKCSone-format oracle to validate received inputs. 
Both the controller and TCP server were implemented in Python3. 
Additionally, we created test harnesses in C/C++ for each test subject.

While prior work \cite{kuhn2008BleichenbacherVariants,chau2019analyzing}, including our own \cite{yahyazadeh2021morpheus}, 
examined several \PKCSone signature verification libraries 
and discovered vulnerabilities capable of leading to signature forgery \cite{finney}, 
our study in this paper is prompted by bugs (\eg, memory-corruption) guarded by initial input validation \cite{yahyazadeh2021morpheus,brubaker2014frankencerts}. To uncover these bugs, a fuzzer needs 
to generate semantically valid inputs.
Therefore, 
this study addresses two research questions: 
evaluating each fuzzer's effectiveness in generating valid inputs for \PKCSone libraries and 
assessing the diversity of inputs produced. 
Through meticulous experimentation and metrics such as the percentage of valid inputs, 
edit distance, and normalized longest common subsequence, 
the research reveals nuanced performance differences. 
\peach \cite{peachfuzz} was found to be the least effective in generating valid inputs among grammar-based black-box fuzzers, 
while \nautilus excelled among grey-box fuzzers. 
\aflsmart \cite{pham2019smart}, despite relying on input grammars, produced fewer valid inputs
compared to other \afl variants. \isla \cite{steinhofel2022input} stood out for 
producing the most diverse inputs but with the lowest throughput (1.3 inputs per second).

Our investigation was propelled by the understanding that if a fuzzer encounters difficulties 
in generating inputs for \PKCSone, similar challenges are likely to arise with 
context-sensitive inputs for other protocols, such as TCP and DNS. 
Our findings  underscore the limitations of existing fuzzers in handling context-sensitive inputs. 
We anticipate that these results will motivate the community to pursue the development of 
improved general-purpose fuzzers.
Our tool is available at \texttt{https://github.com/syne-lab/fuzz-eval}.

\section{Preliminaries}
\label{sec:background}

This section briefly covers fuzzing basics 
and \PKCSone standard for RSA signature generation and 
verification. 

\subsection{Fuzzing}\label{subsec:fuzzing}

Fuzzing is a widely-used software testing approach \cite{fuzzing2021} 
that systematically feeds abnormal inputs into a program to uncover crashes or unexpected behaviors, often related to security vulnerabilities
and bugs within the program under test. 

Fuzzers are classified into three types: white-box, grey-box, and black-box. White-box fuzzers, 
like SAGE  \cite{godefroid2012sage}, rely on detailed program understanding, often from techniques 
such as symbolic execution \cite{cadar2013symbolic}. 
Grey-box fuzzers, such as \afl \cite{afl} and \libfuzzer \cite{libfuzzer}, 
work with partial information like code coverage metrics. 
Black-box fuzzers like \peach \cite{peachfuzz} and \morpheus \cite{yahyazadeh2021morpheus} 
operate without knowledge of the program's internal structure, emphasizing simplicity and speed.

Fuzzers collect execution data from tests to generate new ones. White-box fuzzers prioritize path constraints, grey-box fuzzers use coverage metrics like 
block or branch coverage, while black-box fuzzers rely solely on input mutations due to 
the lack of runtime knowledge of the target program.

Fuzzers employ diverse input generation methods, from basic random mutations 
\cite{swiecki2016honggfuzz, libfuzzer, afl} to advanced grammar-based techniques 
\cite{aschermann2019nautilus,pham2019smart,steinhofel2022input,peachfuzz}. 
Random mutation methods alter existing inputs randomly, while grammar-based approaches 
utilize input structure and relationships to generate inputs.

Overall, fuzzing is potent in revealing software bugs and vulnerabilities, with approaches 
presenting varying trade-offs between program insight and testing efficiency.

\subsection{\PKCSone and RSA Signatures}\label{subsec:PKCS-basics}

\PKCSone \cite{rfc2313}, a cryptographic standard, plays a pivotal role in secure communication protocols 
such as SSL/TLS and IPSec
as well as in software signing and X.509 certificates. It defines formats for RSA encryption and signature schemes, including padding schemes.

To generate a signature for a message, \code{M}, using the RSA signature scheme and \PKCSone padding scheme,
the message is encoded to produce an encoded message, \code{EM}, structured as follows:

\begin{center}
	\vspace*{-0.5em}
\code{EM = 0x00 $\|$ BT $\|$ PS $\|$ 0x00 $\|$ PL}
\end{center}

Here, \code{BT} represents the block type (\code{0x01} for signatures, \code{0x02} for encryption),
\code{PS} denotes a padding string (byte sequence containing \code{0xFF}) for RSA signatures, and \code{PL} represents the payload bytes.
The signature, \code{S}, is generated using the operation $S = \code{EM}^d$ \code{mod} $n$, where $d$ is the RSA private exponent and $n$ is the public modulus.

To ensure the generation of valid \code{EM} structures, it is crucial to maintain 
$|\code{EM}| = |n|$, where $|n|$ denotes the size of the public modulus in bytes.
Additional constraints on the \code{EM} structure include the minimum length of the padding string ($|\code{PS}| \ge 8$) 
and the relationship between \code{PS} and \code{PL} ($|\code{PS}| = |n| - |\code{PL}| - 3$).
These constraints, along with idiosyncrasies among different \code{EM} fields, render \PKCSone 
ideal for evaluating fuzzers' effectiveness in generating context-sensitive inputs.

 \section{Design and Implementation}

We will now explain the design and implementation of our 
platform, FuzzEval.

\subsection{Design of FuzzEval}\label{subsec:workflow}

We have designed FuzzEval, a comprehensive platform for assessing fuzzer effectiveness 
in generating context-sensitive inputs. 
FuzzEval ensures consistent testing environments across various fuzz campaigns via customizable settings. 
It comprises three core components: Controller, Validator, and Oracle. 
The controller orchestrates fuzz campaigns, while the validator receives fuzzer-generated inputs, verifies their validity using the oracle based on the 
\PKCSone specification \cite{rfc2313}, and logs results. 
FuzzEval seamlessly integrates several fuzzers, test harnesses, and test subjects.

\noindent\textbf{Controller.}
The controller, the cornerstone of FuzzEval, operates based on a configuration file 
containing crucial campaign details: paths to fuzzer and test-subject executables, the fuzzing command, and 
parameters like duration, seed directory, and validator server port. This file provides precise control over the fuzzing environment.

After parsing the configuration, the controller sets up the validator server on the designated port and 
launches the fuzzing campaign for the $\langle$fuzzer, test subject$\rangle$ pair. 
It monitors the campaign's progress until completion, 
then gracefully shuts down the validator server. The Controller can handle multiple campaigns simultaneously.

\begin{figure}[!t]
	\centering
	\includegraphics[width=0.8\columnwidth]{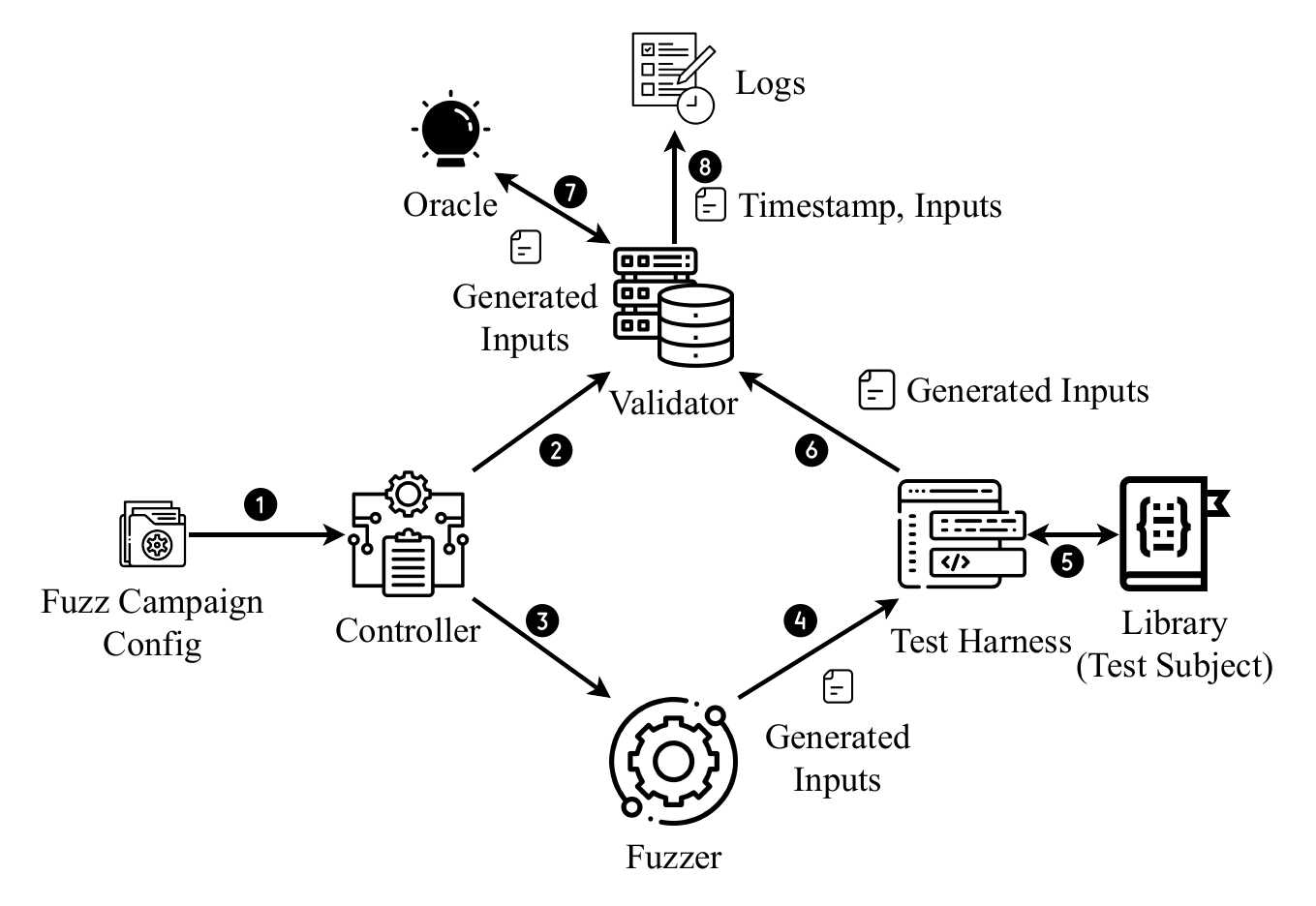}
	\vspace*{-0.2in}
	\caption{Workflow of FuzzEval}
	\label{fig:workflow}
	\vspace*{-0.2in}
\end{figure}

\noindent\textbf{Validator.}
The validator, a locally hosted TCP server, receives fuzzer-generated inputs from test harnesses. 
Its main role is to validate these inputs by consulting the oracle. Each input is sent to the oracle, 
and the validator then records the oracle's response, along with the original input and timestamp, 
in a designated file, which will be analyzed later for assessing the fuzzer's performance.

\noindent\textbf{Oracle.}
The oracle validates fuzzer-generated inputs' integrity according to the \PKCSone 
specification \cite{rfc2313}. 
While the payload (\code{PL}) usually includes a valid cryptographic hash of the signed message, 
we do not verify its validity against specific hash function families. 
Instead, we treat it as a byte sequence. 
Our focus lies in confirming if each generated input (\ie, \code{EM}) matches 
the predefined public modulus's length and adheres to the structural properties outlined in 
$\S$~\ref{subsec:PKCS-basics}.

\noindent\textbf{Fuzzer, Test Harness, and Test Subject.}
Instead of directly fuzzing a test subject (a library), 
we focus on fuzzing a test harness that uses the library for RSA signature verification. 
The fuzzer runs the test harness, which expects a \PKCSone formatted byte array (\texttt{EM} structure). 
The harness generates an RSA signature using this encoded message (\texttt{EM}). 
The RSA components (public modulus, public exponent, and private exponent) remain constant across all test subjects. 
After generating the signature, the harness executes the test subject by calling its signature verification function. 
Once the function returns, the harness forwards the original \texttt{EM} structure to FuzzEval's validator to check if it conforms to \PKCSone. These steps are repeated with each new \texttt{EM} generated by the fuzzer. 
See Appendix \ref{appendix:harness} for a sample test harness.

\noindent\textbf{Workflow.}
Figure \ref{fig:workflow} depicts FuzzEval's operation. 
In summary, the controller takes in the fuzz campaign configuration (\circled{1}), launches the validator server (\circled{2}), initiates the $\langle$fuzzer, test-subject$\rangle$ pair fuzzing campaign (\circled{3}), where the fuzzer provides inputs to the test harness (\circled{4}). The test harness processes these inputs using the test subject library (\circled{5}) and sends them to the validator (\circled{6}). The validator then consults the oracle to validate the input (\circled{7}) and logs the input, along with the oracle's response and timestamp (\circled{8}). 
Upon campaign completion, the controller gracefully shuts down the validator server.

\subsection{Implementation}\label{subsec:implementation}
FuzzEval is primarily implemented in Python 3. 
We additionally developed test harnesses for each \PKCSone library in C or C++. 
The controller, written in Python 3, processes a TOML configuration file. 
For TCP functionality, the validator and controller utilize the Twisted library in Python 3. 
The critical oracle component is a Python 3 function. Log files store fuzzer-generated inputs, oracle's decisions, 
and library responses. 
For consistency and reproducibility across multiple fuzzing campaigns, we employ a Linux-container-based virtualized environment (using Docker), dedicating each container to a single campaign.

 \section{Evaluation}
\label{sec:eval}

Unlike prior work \cite{kuhn2008BleichenbacherVariants,chau2019analyzing,finney,yahyazadeh2021morpheus} 
focused on discovering vulnerabilities in \pkcs libraries that could lead to signature forgery, 
our study is propelled 
by vulnerabilities (\eg, memory corruption) hidden behind initial input validation (\ie, sanity checks). 
Since a fuzzer capable of generating valid inputs can uncover such bugs \cite{yahyazadeh2021morpheus}, 
we focus on assessing several existing fuzzers' performance in generating valid and 
diverse inputs for complex implementations like 
\pkcs libraries. We leave the assessment of the fuzzers' effectiveness in discovering 
new vulnerabilities with valid inputs for future work.

\noindent\textbf{Research Questions and Metrics.}
We seek to answer the following research questions:
\begin{itemize}[leftmargin=*,topsep=0pt]
	\item \textbf{RQ1:} How effective are the fuzzers in generating valid inputs for \PKCSone libraries?
	\item \textbf{RQ2:} How diverse are the generated inputs? \end{itemize}

To address RQ1, we measure the percentage of valid inputs generated by each fuzzer 
for its respective test subject over time. A higher percentage of valid inputs indicates greater effectiveness in generating valid inputs.

To tackle RQ2, we employ two metrics borrowed from \cite{ouyang2023llm}: 
(1) \editdist, the edit distance between two input strings $A$ and $B$, and (2) \lcs, the normalized longest common subsequence between two input strings $A$ and $B$. 
\editdist quantifies the dissimilarity between input strings, while \lcs measures their similarity. 
Higher edit distance and lower normalized longest common subsequence both independently signify greater input diversity.
Appendix \ref{appendix:rq2-metrics} shows the details of \editdist and \lcs.

Given the computational expenses associated with calculating both \editdist and \lcs for pairwise inputs, 
which can number in the hundreds of thousands, we opted to sample 100 inputs per fuzzer for each test subject 
when determining these metrics. Following the methodology outlined in \cite{samplesize}, we selected a 
sample size of 100. Subsequently, we computed the average across the sampled inputs. 
To mitigate the effects of randomness, we repeated this process 10 times and reported the resulting average.

\subsection{Experimental Setup}

\noindent\textbf{Fuzzers.} 
Our objective was to encompass a wide range of fuzzers with diverse characteristics, spanning from black-box to grey-box approaches, employing various techniques such as input grammars or seed inputs, and differing in their reliance on human assistance or access to source code. We evaluated eleven fuzzers, including established ones like \aflplusplus, 
\libfuzzer, and \honggfuzz, alongside newer additions like \darwin, \isla and \sgfuzz. The distinctive features of these fuzzers are outlined in Table \ref{tab:fuzzertable}.

\afl, \aflplusplus, \libfuzzer, and \honggfuzz operate as grey-box fuzzers and typically demonstrate improved performance when provided with valid seed inputs. \darwin and \sgfuzz represent recent advancements in grey-box fuzzing techniques. In contrast, \peach and \isla 
function as black-box fuzzers, necessitating input format specifications in their respective languages. 
Note that
{\isla is technically an input generator for a fuzzer \cite{steinhofel2022input}.
Similarly, \morpheus, another black-box fuzzer, is specialized for testing \PKCSone implementations and 
internally embeds the \PKCSone specifications. \nautilus and \aflsmart are grey-box fuzzers, although they rely on input specifications. While \nautilus operates without seed inputs, \aflsmart benefits from having valid seed inputs.

\aflsmart, \darwin, and \aflplusplus share similarities with \afl in terms of instrumentation and input generation. 
In \afl, random inputs are generated by mutating previously generated inputs through \textit{deterministic} stages 
(targeted mutations) and the havoc stage (randomness). 
All of these fuzzers allow selective enabling or disabling of these stages, except for \aflsmart. 
In our evaluation, we assessed these fuzzers with both deterministic fuzzing mode enabled and disabled 
whenever applicable, denoted by an asterisk (*) like \afl*.

\noindent\textbf{Test Subjects (Libraries).} 
To comprehensively assess and evaluate the proficiency of the fuzzers in generating inputs, 
we utilized 13 libraries implementing \PKCSone signature verification. These libraries were selected based 
on their widespread usage and implemented in C/C++, as detailed in Table \ref{tab:testsubjecttable}.
Instead of directly fuzzing a test library,
we focused on fuzzing a test harness that employs the library for RSA signature 
verification. The harness takes in 
a \PKCSone formatted byte array (an \code{EM} structure, see $\S$~\ref{subsec:PKCS-basics}), 
generates an RSA signature using \code{EM}, predefined private key $(n,d)$, and a message (\eg, \code{"hello world!"}),
and utilizes the library to verify the RSA signature. Figure~\ref{fig:harness} illustrates a sample harness.
Note that the harness sends the original \texttt{EM} structure to FuzzEval's validator 
to check the validity of the \texttt{EM} structure, 
not the signature or the outcomes of the library's cryptographic operations.

\noindent\textbf{Computing Setup.} 
Each fuzzer underwent a 10-hour testing period on each library, with five repeated experiments. The evaluation was conducted on a server with a 128-core 2.0 GHz Intel Xeon Gold 6338 CPU and 256GB RAM. Each fuzzing campaign ran in a dedicated Ubuntu 20.04 Docker container, allocated 8GB of RAM and 3 cores. To reduce costs, we chose a 10-hour testing period, resulting in a total of 9,100 computing hours and an estimated cost of \$1,800 USD, based on Amazon EC2 \texttt{c7a.xlarge} instance pricing.

\subsection{Results}

\noindent\textbf{RQ1.} Table \ref{tab:validity-short} shows the percentage of valid inputs 
generated by some fuzzers for a few test subjects (Table \ref{tab:summary} shows the numerical values for each $\langle$fuzzer, test-subject$\rangle$ pair).
For reference, Figure~\ref{fig:full-periodicdata} shows how the rates change over time for each pair.

\isla stands out by consistently generating 100\% valid inputs for all test subjects (not shown in Table \ref{tab:validity-short}).
\isla takes in the input grammar of the \code{EM} structure and employs an SMT solver (\eg, Z3 \cite{z3})
to resolve \code{EM}'s constraints during input generation.
As a result, it invariably generates 100\% valid inputs.
Since \isla does not require the program under test to generate inputs, 
its results were not included in the table and figure. 

Following \isla, \morpheus emerges as the next most effective fuzzer, producing 56.69\% 
valid inputs for all test subjects (not shown in Table \ref{tab:validity-short}). 
As a black-box fuzzer, \morpheus operates based on \PKCSone specifications without direct access to the tested library. 
It employs a set of predefined mutations during testing, 
resulting in a limited number of valid inputs compared to \isla.
The results of \morpheus is also not shown in the table.

Conversely, \peach, another black-box fuzzer, proves to be the least effective among its peers, 
generating only 17.79\% valid inputs on average. 
Despite relying on input specifications, \peach only permits a 
context-free grammar of the \code{EM} structure to be supplied in its language.
This limitation accounts for its poor performance, as \peach lacks awareness of the necessary constraints of EM.

Among the grey-box fuzzers, \nautilus generates the most valid inputs (21.92\%) overall. However, it 
shows lower effectiveness for \texttt{strongswan}, \texttt{mbedtls}, \texttt{cyrpto++} and \texttt{gnutls}. 
Among the \afl-based fuzzers, \afl, \aflplusplus,
and \darwin generated significant valid inputs when running with deterministic 
fuzzing mode enabled (denoted by an asterisk, \eg, \afl{*}). 
Disabling this mode results in a notable decrease in valid inputs. 
For example, when \afl fuzzed axtls, the percentage of valid inputs dropped from 24.58\% to 0.51\%.

\begin{table*}[!t]
	\footnotesize
	\centering
	\caption{The percentage of valid inputs 
	generated by some fuzzers across a few test subjects. 
A higher value indicates greater effectiveness of the fuzzer in generating valid inputs.
	Standard deviations of the rates are presented in parentheses.
}
	\label{tab:validity-short}
	\vspace*{-1em}
	\begin{tabular}{|p{1.3cm}|c|c|c|c|c|c|c|}
		\hline
		\textbf{Test-subject} & \textbf{\afl} & \textbf{\afl*} & \textbf{\aflplusplus} & \textbf{\darwin} & \textbf{\aflsmart} & \textbf{\nautilus} & \textbf{\peach} \\ \hline
		{axtls}                 &  0.51 (0.14)  &  24.58 (1.83)  &      1.59 (0.24)      &   1.80 (0.55)    &    0.58 (0.23)     &    18.94 (2.43)    &  17.54 (0.76)   \\ \hline
		{crypto++}              &  0.08 (0.02)  &  7.98 (3.01)   &      0.50 (0.26)      &   1.28 (0.64)    &    0.19 (0.12)     &    0.09 (0.01)     &  17.92 (0.03)   \\ \hline
		{gnutls}                &  1.00 (0.11)  &  4.51 (0.13)   &      1.83 (0.09)      &   1.83 (0.63)    &    3.45 (0.02)     &    0.18 (0.00)     &  17.85 (0.07)   \\ \hline
		{mbedtls}               &  0.12 (0.02)  &  4.15 (0.05)   &      1.33 (0.01)      &   0.37 (0.19)    &    0.35 (0.05)     &    0.18 (0.00)     &  17.89 (0.05)   \\ \hline
		{strongswan}            &  0.19 (0.08)  &  9.16 (2.41)   &      1.55 (0.68)      &   2.82 (0.95)    &    0.29 (0.15)     &    0.13 (0.04)     &  17.88 (0.02)   \\ \hline
		\textit{omitted}               &    -- (--)    &    -- (--)     &        -- (--)        &     -- (--)      &      -- (--)       &      -- (--)       &     -- (--)     \\ \hline
		\textbf{Average}               &  1.45 (1.77)  & 16.75 (10.58)  &      3.35 (3.04)      &   3.72 (3.73)    &    1.78 (1.97)     &   21.92(17.75)    &  17.79 (0.25)   \\ \hline
	\end{tabular}
\vspace*{-0.2in}
\end{table*}

Although \aflsmart requires input specifications, it proved less effective compared to other AFL-based fuzzers. 
For example, while \afl{}* generated 24.58\% valid inputs for axtls, \aflsmart only produced 0.58\% 
valid inputs for the same target. This disparity can be attributed to \aflsmart's heavy reliance 
on havoc-based mutations, which drastically alter even valid seed inputs. 
Unfortunately, \aflsmart lacks an option to disable these mutations. 
The remaining fuzzers generated an insignificant number of valid inputs across all test subjects.

We observed an overall trend: black-box fuzzers relying on input specifications, such as \morpheus, \isla, and \peach, tend to outperform grey-box fuzzers in generating semantically valid inputs. However, code-coverage feedback can derail grey-box fuzzers, even if they start with good seed inputs 
(\eg, \aflplusplus), input specifications (\eg, \nautilus), or both (\eg, \aflsmart). This is expected since
grey-box fuzzers focus more on the test-subject's code coverage instead of generating semantically valid inputs.

\noindent\textit{Throughput.} We measured the throughput of each fuzzer, representing
the number of inputs (valid or invalid) generated per second across all test-subjects.
Fuzzers generating numerous valid inputs, such as \isla and \morpheus, exhibit slower throughput.
For example,  \isla generated 100\% valid inputs but at the lowest 
throughput of 1.30 inputs per second due to its use of an SMT solver during input generation. 
Conversely, fuzzers producing many invalid inputs tend to 
generate inputs at a higher speed, 
prioritizing speed over context-sensitive constraints (see Figure \ref{fig:throughput}).

\noindent\textbf{RQ2.} 
We measured the diversity of the generated inputs using \editdist (edit distance)
and \lcs (normalized longest common subsequence). For reference, 
the measured \editdist and \lcs for each $\langle$fuzzer, test-subject$\rangle$ pair 
are shown in Tables \ref{tab:editdist} and \ref{tab:lcs}. 

Among black-box fuzzers, \peach produced the least diverse inputs for all test subjects (\editdist: 22.14, 
\lcs: 0.98), 
Morpheus outperformed 
\peach (\editdist: 49.73, \lcs: 0.94).
We left the reporting of \editdist and \lcs for \isla since
\isla produced 100\% valid inputs (\ie, \code{EM} structures) conforming to \PKCSone, with little to no diversity.
Conversely,
among grey-box fuzzers, \sgfuzz generated the least diverse inputs (\editdist: 254.29, \lcs: 0.45), 
while \darwin produced the most diverse ones (\editdist: 4356.88, \lcs: 0.48).

These diversity metrics revealed a trend: 
Fuzzers generating many valid inputs tend to produce inputs 
with lower edit distance (\editdist in the hundreds or less), 
while those generating many invalid inputs produce inputs with higher edit distance (\editdist in the thousands). 
This aligns with expectations, as valid inputs must adhere closely to specific formats, 
while invalid inputs have more variability.
A similar trend was observed with \lcs, where valid inputs exhibit higher \lcs values as they share
more portions, while invalid inputs show lower \lcs values 
due to less shared content.

\begin{figure}[!t]
	\centering
	\includegraphics[width=0.8\linewidth]{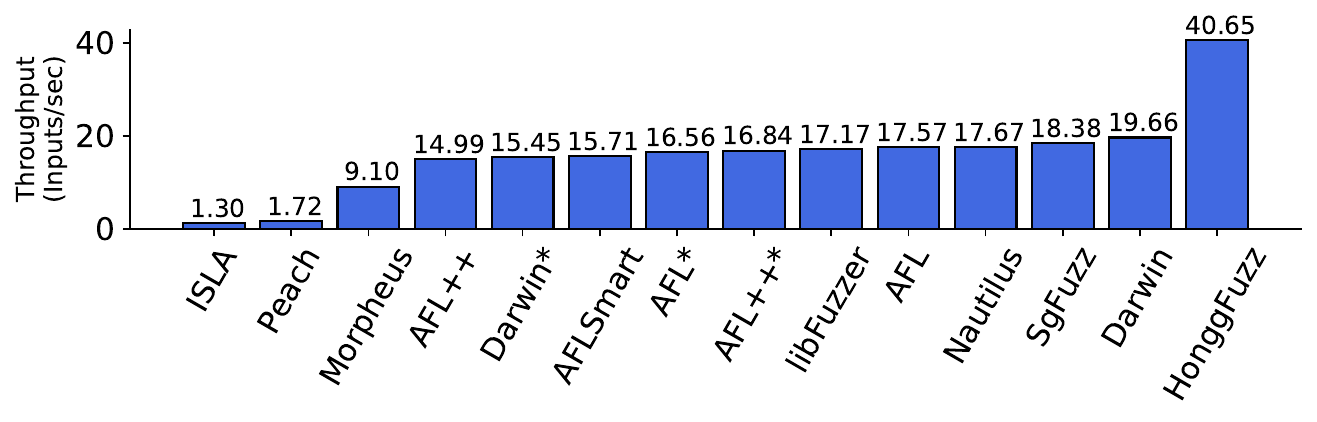}
	\vspace*{-0.2in}
	\caption{The average throughput of each fuzzer}
	\label{fig:throughput}
	\vspace*{-0.2in}
\end{figure} \section{Takeaways}
\label{sec:lessons}

\noindent\textbf{Domain-specific vs. General-purpose Tools.}
While domain-specific fuzzers like \morpheus are expected to excel, 
general-purpose tools like \isla can surprisingly outperform \morpheus 
due to their support for context-sensitive constraints and utilization of an SMT solver. 
However, \isla's superiority comes at the expense of lower throughput, 
making it less suitable for fuzzing scenarios prioritizing speed.

\noindent\textbf{Input Validity vs. Diversity.}
Fuzzers generating valid inputs tend to exhibit less diversity in terms of edit distance (\editdist)
and normalized longest common subsequence (\lcs). 
This is logical, as valid inputs must adhere to specific formats, 
limiting variability. Conversely, fuzzers generating more invalid inputs 
prioritize producing significantly different inputs, 
emphasizing either broader input testing or code coverage.

\noindent\textbf{Deterministic fuzzing mode.}
\afl-like grey-box fuzzers with deterministic fuzzing mode enabled can outperform 
some grammar-based fuzzers like \peach when provided with valid seed inputs. 
This enables them to generate multiple valid inputs by mutating less critical parts 
of the input structure, such as the payload (\code{PL}) of the \PKCSone structure. 
Consequently, these fuzzers uncover new paths of the test-subject and 
continue mutating corresponding valid inputs. 
Conversely, disabling deterministic fuzzing mode often leads to input structure breakage, resulting in more invalid inputs.

\noindent\textbf{Context-free input grammars.} 
Fuzzers dependent on context-free input specifications, like \nautilus, \aflsmart, and \peach,
struggle to address context-sensitive constraints, leading to numerous invalid inputs.

\noindent\textbf{Influence of Test-subjects.}
Grey-box fuzzers, regardless of reliance on input specifications, are influenced by each test subject. 
The rate of valid input generation varies with both the test subject and time, as shown in Figure~\ref{fig:full-periodicdata}.

\noindent\textbf{Threat to Validity.} 
Testing 11 fuzzers across 13 libraries for ten hours may not fully capture the fuzzers' 
potential for generating valid inputs, impacting overall generalizability.

 \section{Related Work}\label{sec:relatedwork}

\noindent
\textbf{Fuzzer Evaluation.} 
LAVA \cite{lava} introduces crashing bugs for testing but may produce bugs that 
differ from real-world vulnerabilities \cite{klees2018evaluating, bundt2021evaluating, aschermann2019redqueen}. 
Similarly, Evil Coder \cite{pewny2016evilcoder} employs data flow analysis to insert bugs. 
However, our focus centers on comparing fuzzers based on their capacity to generate intricate, context-sensitive inputs. 
Unlike benchmarks such as Cyber Grand Challenge (CGC) binaries \cite{darpaCGC}, 
which often simulate real programs and may introduce biases \cite{klees2018evaluating}, 
our work evaluates widely adopted cryptographic libraries implementing the \PKCSone signature scheme, 
aligning assessments with real-world applications. 
While we acknowledge that our selection of 13 libraries may not constitute 
a comprehensive benchmark suite, we recognize the ongoing challenge of deriving such a suite.

Magma \cite{hazimeh2020magma} integrates real-world vulnerabilities into software projects, assessing triggering inputs but requires manual effort for version updates and may introduce biases. 
In contrast, FuzzEval avoids code-base modification.

UniFuzz \cite{li2021unifuzz} and Fuzzbench \cite{metzman2021fuzzbench} employ real programs 
for fuzzing benchmarks. UniFuzz lacks ground-truth data, posing a challenge for 
fuzzer comparison, whereas Fuzzbench prioritizes coverage. 
Our study stands out by uniquely focusing on comparing fuzzers based on their input generation proficiency, 
unlike existing research focused on coverage or bug metrics.

Metrics based on coverage or bugs may be unreliable \cite{bohme2022reliability}, 
as they might not capture a fuzzer's ability to explore deep program states. 
For example, when fuzzing a program handling structured inputs like XML, a fuzzer might 
uncover parsing bugs but miss core logic issues. Semantically valid inputs have the potential to reach deep program 
states, and therefore, 
our study prioritizes evaluating fuzzers based on their capability to create semantically valid inputs. 

\noindent\textbf{Evaluation of \PKCSone.} 
Despite theoretical security claims \cite{pkcsproof}, widely used implementations of the \PKCSone signature scheme 
have vulnerabilities, potentially leading to signature forgery. 
Automated evaluations discovered flaws in several implementations using symbolic execution \cite{chau2019analyzing} and 
a domain-specific fuzzer \cite{yahyazadeh2021morpheus}. 
Our study aims to assess eleven fuzzers' performance in generating intricate, context-sensitive inputs, 
with a focus on the \PKCSone signature scheme.

 \section{Conclusion and Future Work}

Our comprehensive evaluation of eleven state-of-the-art fuzzers has shed light on their efficacy in generating 
semantically valid inputs for testing cryptographic standards, \eg \PKCSone. 
Through testing across thirteen implementations, we uncovered nuanced performance differences 
among the fuzzers in terms of both validity and diversity of generated inputs. 
Our findings highlight the persistent challenges and limitations faced by existing fuzzers 
in handling context-sensitive inputs. Future work will focus on evaluating how the generated valid inputs influence 
other performance metrics (\eg, code-coverage, new bugs) of fuzzers.

\section*{Acknowledgment}
This research was supported, in part, by the NSF award CNS-2339350 and a gift from Google Research.
 
\bibliographystyle{IEEEtran}
\bibliography{ref}

\appendix

\section{Test Subjects: \PKCSone Libraries}\label{appendix:fuzzersandtestsubjects}

\begin{table}[!t]
	\centering
	\scriptsize
	\caption{List of test subjects and their versions.}
\begin{tabular}{|c|c|}
	\hline
	\textbf{Library Name} & \textbf{Version / Git-commit Id} \\ \hline
	axtls                 & v 2.1.5                          \\ \hline
	botan                 & v 2.17.3                         \\ \hline
	crypto++              & v 8.5.0                          \\ \hline
	gnutls                & v 3.6.15                         \\ \hline
	hostapd               & v 2.9                            \\ \hline
	libtomcrypt           & v 1.18.2                         \\ \hline
	matrixssl             & v 4.3.0                          \\ \hline
	mbedtls             & v 2.26.0                         \\ \hline
	openssl3              & GC - c74188e                     \\ \hline
	openssl1              & v 1.1.1k                         \\ \hline
	openswan              & v 3.0.0                          \\ \hline
	strongswan            & v 5.9.2                          \\ \hline
	wpasupplicant         & v 2.9                            \\ \hline
	\end{tabular}

	\label{tab:testsubjecttable}
\end{table} 
Table \ref{tab:testsubjecttable} lists the test subjects used in our evaluation. The
Version/Git-Commit Id column lists the version number or the Git commit hash of the test
subject.

\section{Fuzzers}\label{appendix:fuzzers}

Table \ref{tab:fuzzertable} lists the fuzzers used in our evaluation.

\begin{table*}[]
\footnotesize
	\caption{List of compared fuzzers and their characteristics. \halfcirc: Greybox fuzzer, \fullcirc: Blackbox fuzzer,
	 \cmark: Required / Supported, \xmark: Not Required / Not Supported, \smark: Works w/o source code.}
\begin{center}

	\begin{tabular}{|l|c|c|c|c|c|c|}
	\hline
	\textbf{Fuzzer}               &    \afl    &       \aflplusplus        &      \aflsmart       &    \libfuzzer    &        \sgfuzz        &          \darwin          \\ \hline
	\textbf{Citation}             & \cite{afl} & \cite{AFLplusplus-Woot20} & \cite{pham2019smart} & \cite{libfuzzer} & \cite{ba2022stateful} & \cite{jauernig2022darwin} \\ \hline
	\textbf{Release Year}         &    2013    &           2023            &         2019         &       2022       &         2022          &           2023            \\ \hline
	\textbf{Version/Commit}       &   v2.57b   &          v4.07a           &       4286ae47       &     v14.0.0      &       00dbbd70        &         dc51fc4e          \\ \hline
	\textbf{Commit Year}          &    2020    &           2023            &         2022         &       2022       &         2023          &           2023            \\ \hline
	\textbf{Fuzzer Category}      & \halfcirc  &         \halfcirc         &      \halfcirc       &    \halfcirc     &       \halfcirc       &         \halfcirc         \\ \hline
	\textbf{Human Assistance}     &   \xmark   &          \xmark           &        \cmark        &      \xmark      &        \xmark         &          \xmark           \\ \hline
	\textbf{Source Code}          &   \smark   &          \smark           &        \smark        &      \cmark      &        \cmark         &          \cmark           \\ \hline
	\textbf{Good Seed Corpus}     &   \xmark   &          \xmark           &        \cmark        &      \xmark      &        \xmark         &          \xmark           \\ \hline
	\textbf{Format Specification} &   \xmark   &          \xmark           &        \cmark        &      \xmark      &        \xmark         &          \xmark           \\ \hline
	\textbf{Large-scale Mutation} &   \xmark   &          \cmark           &        \cmark        &      \xmark      &        \cmark         &          \cmark           \\ \hline
\end{tabular}
\medskip

	\begin{tabular}{|l|c|c|c|c|c|}
	\hline
	\textbf{Fuzzer}               &           \nautilus           &         \honggfuzz          &           \morpheus           &           \isla            &      \peach      \\ \hline
	\textbf{Citation}             & \cite{aschermann2019nautilus} & \cite{swiecki2016honggfuzz} & \cite{yahyazadeh2021morpheus} & \cite{steinhofel2022input} & \cite{peachfuzz} \\ \hline
	\textbf{Release Year}         &             2019              &            2015             &             2022              &            2022            &       2013       \\ \hline
	\textbf{Version/Commit}       &           4cfeb26f            &          3a8f2ae4           &           626d2e81            &          48af902           &     v3.0.202     \\ \hline
	\textbf{Commit Year}          &             2023              &            2023             &             2022              &            2023            &       2013       \\ \hline
	\textbf{Fuzzer Category}      &           \halfcirc           &          \halfcirc          &           \fullcirc           &         \fullcirc          &    \fullcirc     \\ \hline
	\textbf{Human Assistance}     &            \cmark             &           \xmark            &            \cmark             &           \cmark           &      \cmark      \\ \hline
	\textbf{Source Code}          &            \cmark             &           \smark            &            \xmark             &           \xmark           &      \xmark      \\ \hline
	\textbf{Good Seed Corpus}     &            \xmark             &           \xmark            &            \xmark             &           \xmark           &      \xmark      \\ \hline
	\textbf{Format Specification} &            \cmark             &           \xmark            &            \cmark             &           \cmark           &      \cmark      \\ \hline
	\textbf{Large-scale Mutation} &            \cmark             &           \xmark            &            \cmark             &           \cmark           &      \xmark      \\ \hline
\end{tabular}

\end{center}

	\label{tab:fuzzertable}
\end{table*}

\section{Diversity Metrics}\label{appendix:rq2-metrics}

\begin{figure}[!t]
	\centering
	\begin{scriptsize}
		\[
\editdist[i, j] = \label{eq:editdist} \tag{Eq 1}
		\begin{cases}
		0 & \text{if } i = 0 \text{ or } j = 0 \\
		\editdist[i-1, j-1] & \text{if } A[i] = B[j] \\
		1 + \min(\editdist[i-1, j], \editdist[i, j-1], \\\qquad \editdist[i-1, j-1]) & \text{otherwise}
		\end{cases}	
		\]
		\[
\lcs = \frac{len(\lcsfunc(A, B))}{len(A)} \label{eq:lcs} \tag{Eq 2}	
		\]
	\end{scriptsize}
	\vspace*{-0.2in}
	\caption{Diversity metrics for RQ2}
	\label{fig:metrics}
\end{figure}

\ref{eq:editdist} of Figure~\ref{fig:metrics} illustrates the calculation of \editdist, representing the edit distance between 
the first $i$ characters of $A$ and the first $j$ characters of $B$.
Conversely, \ref{eq:lcs} outlines the computation of \lcs, 
with $A$ as the reference string and $B$ as the string under comparison. Here, $\lcsfunc(A,B)$ 
denotes the longest common subsequence shared between $A$ and $B$.

\section{A Sample Harness Program}\label{appendix:harness}

\begin{figure*} \centering
\caption{A sample harness program for test subject \code{Botan}.}
\label{fig:harness}
\noindent\begin{minipage}{.5\linewidth}
\begin{lstlisting}[language=C++, basicstyle=\ttfamily\scriptsize]
#include <botan/rsa.h>
/* Omitted other header files */

uint8_t *datahex(char *string, size_t &dlength) {
  /* Omitted. Convert a string of
   * hex characters to byte array
   */}
unsigned char *read_file_contents(char *filename, int file_size) {
  /* Omitted. This function simply reads 
   * the bytes from the given filename.
   */}
char hexmap[] = {'0', '1', '2', '3',
                 '4', '5', '6', '7',
                 '8', '9', 'a', 'b', 
                 'c', 'd', 'e', 'f'};
std::string hexStr2(unsigned char *data, int len) {
    /* Omitted. Converts the byte array
     * to a hex string */}
sig_atomic_t signaled = 0;
int mod_len = 256;
string hexRep;
int status, valread, client_fd;
struct sockaddr_in serv_addr;
void my_handler(int signum) {
  if ((status = connect(client_fd, (struct sockaddr *)&serv_addr,
                        sizeof(serv_addr))) < 0) {
    printf("\nConnection Failed \n"); }
  hexRep += ",";
  hexRep += "-1"; // indicates that the library crashed!!
  send(client_fd, hexRep.c_str(), hexRep.size(), 0);
  close(client_fd);
  exit(0); }
void register_custom_handler() {
  /* Omitted. Register signals like SIGSEGV 
   * and SIGABRT to call my_handler()  */ }
void unregister_custom_handler() {
	/* Omitted. Reset signals back to normal. */ }
int main(int argc, char **argv) {
  // Step 0 >> Reigster for handling signals
  register_custom_handler();
  // Step 1 >> Take a file name as an input, read the file in binary mode
  if (argc != 3) {
    printf("usage: return 0; }
  if ((client_fd = socket(AF_INET, SOCK_STREAM, 0)) < 0) {
    printf("\n Socket creation error \n");
    return -1; }

\end{lstlisting}
\end{minipage}
\noindent\begin{minipage}{.5\linewidth}
\begin{lstlisting}[language=C++, basicstyle=\ttfamily\scriptsize]
  /* Omitted. This block simply reads 
  * the port from stdin and connects to it
  */
  char *filename;
  struct stat filestatus;
  int file_size;
  unsigned char *file_contents;
  filename = argv[1];
  /* ...  Omitted ... */
  file_contents = read_file_contents(filename, file_size);
  if (file_contents == NULL) {
    printf("--------------\n");
    return 1; }
  // Step 2 >> Convert the binary file contents to hex string (i.e., the input EM)
  hexRep = hexStr2(file_contents,file_size);
  // Step 3 >> Compute the signature using python script
  /* ... Omitted ... */
  callfunc = PyObject_CallObject(func,args);
  string sig_str = std::string(PyUnicode_AsUTF8(callfunc));
  Py_Finalize();
  std::string msg("hello world!");
  std::vector<uint8_t> data(msg.data(), msg.data() + msg.length());
  Botan::BigInt n(/* ... Omitted ... */), e(3);
  Botan::RSA_PublicKey pubk(n, e);
  // Step 4 >> verify signature using the library
  Botan::PK_Verifier verifier(pubk, "EMSA3(SHA-256)");
  size_t sigLen;
  uint8_t *sig = datahex((char *)sig_str.c_str(),sigLen);
  std::vector<uint8_t> sigv(sig, sig + sigLen);
  int vv = verifier.verify_message(data, sigv);
  // Step 5 >> unregister custom
  unregister_custom_handler()
  // Step 6 >> Send the input EM to the FuzzEval's validator
  if ((status = connect(client_fd, (struct sockaddr *)&serv_addr, sizeof(serv_addr)))<0) {
    printf("\nConnection Failed \n");
    return -1; }
  
  send(client_fd, hexRep.c_str(), hexRep.size(), 0);
  close(client_fd);
  free(file_contents);
  return 0; }
\end{lstlisting}
\end{minipage}
\end{figure*}
 Figure \ref{fig:harness} shows a sample harness program for the test subject \texttt{Botan}.

\section{RQ1 Detailed Results}\label{appendix:rq1}

\begin{figure*}[!t]
	\centering
	\includegraphics[width=0.8\linewidth]{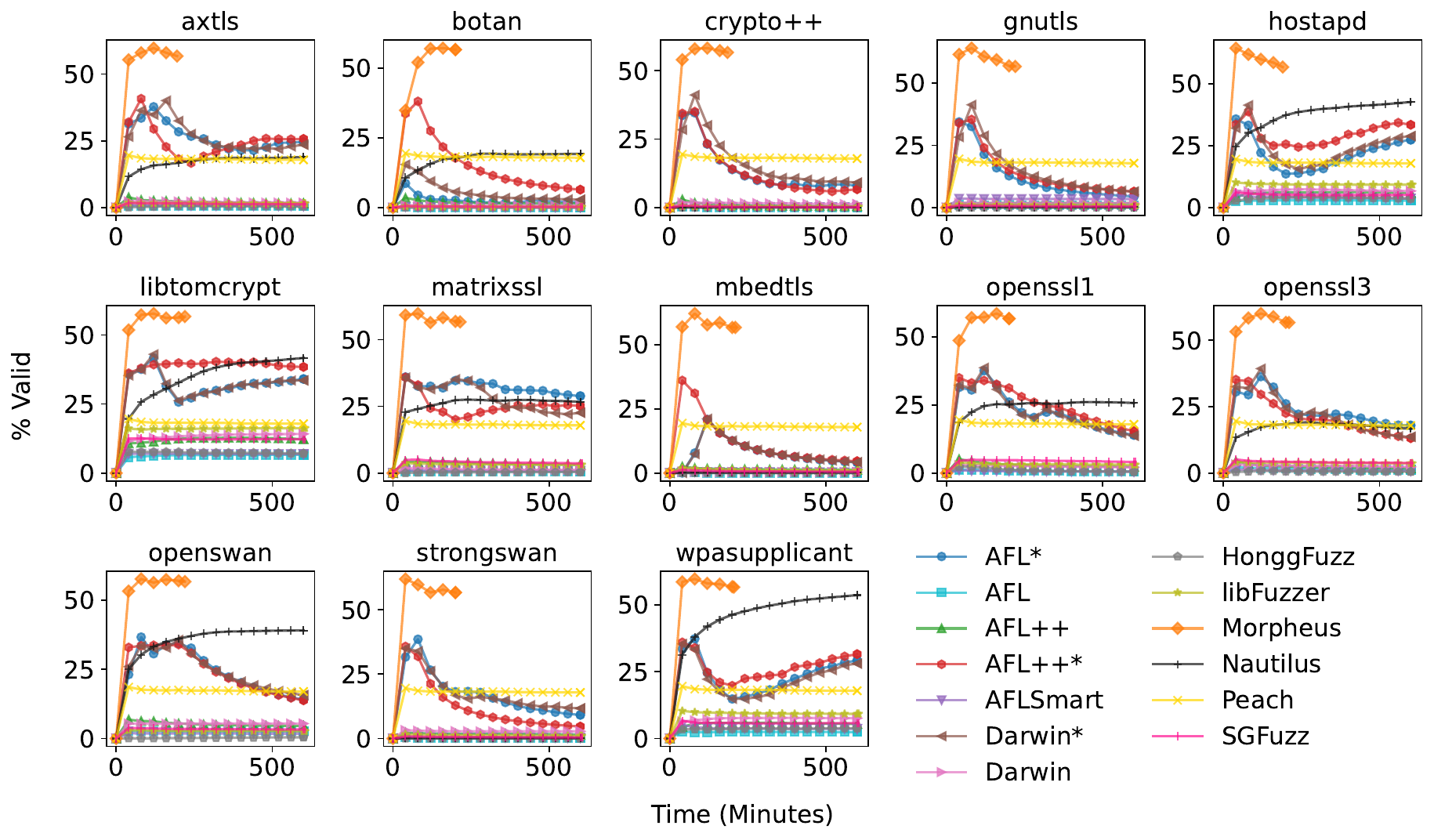}
	\caption{The percentage of valid inputs generated by each fuzzer for each test subject over time (10 hours).
		A fuzzer's name with an asterisk (\eg, AFL*) refers to 
		the campaigns with deterministic fuzzing mode enabled.}
	\label{fig:full-periodicdata}
\end{figure*}

\FIGREF~\ref{fig:full-periodicdata} shows the percentage of valid inputs generated by each fuzzer for each test subject over time (10 hours). Table \ref{tab:summary} shows the numerical values from all 5 runs of 10 hours each. We report the average over 5 runs and the standard deviation of the runs is given in parenthesis.

\begin{table*}[!t]
	\scriptsize
	\centering
	\caption{
	The percentage of valid inputs 
	generated by every fuzzer for each test subject. 
	The reported values represent the averages from five distinct 
	fuzzing campaigns for each $\langle$fuzzer, test-subject$\rangle$ pair.
	A higher value indicates greater effectiveness of the fuzzer in generating valid inputs.
	Standard deviations of the correctness rates are presented in parentheses.
*, like \afl-*, refers to fuzzing campaign with deterministic mode enabled.
	}
	\label{tab:summary}
\begin{tabular}{|l|c|c|c|c|c|c|}
		\hline
		\textbf{Test Subject} & \textbf{\afl} & \textbf{\afl*} & \textbf{\aflplusplus} & \textbf{\aflplusplus*} & \textbf{\darwin} & \textbf{\darwin*} \\ \hline
		\textbf{axtls}         & 0.51 (0.14) & 24.58 (1.83)  & 1.59 (0.24)  & 25.70 (7.11)  & 1.80 (0.55)  & 23.55 (5.07) \\ \hline
		\textbf{botan}         & 0.06 (0.01) & 1.89 (1.60)   & 1.45 (0.02)  & 6.45 (0.26)   & 0.71 (0.36)  & 2.86 (2.20)  \\ \hline
		\textbf{crypto++}      & 0.08 (0.02) & 7.98 (3.01)   & 0.50 (0.26)  & 6.71 (2.32)   & 1.28 (0.64)  & 9.14 (2.90)  \\ \hline
		\textbf{gnutls}        & 1.00 (0.11) & 4.51 (0.13)   & 1.83 (0.09)  & 6.45 (0.28)   & 1.83 (0.63)  & 6.43 (0.20)  \\ \hline
		\textbf{hostapd}       & 2.83 (0.69) & 27.36 (4.33)  & 5.90 (0.75)  & 33.41 (3.83)  & 6.70 (1.48)  & 28.86 (1.91) \\ \hline
		\textbf{libtomcrypt}   & 6.64 (1.69) & 34.07 (0.42)  & 12.27 (2.84) & 38.43 (3.05)  & 14.18 (2.52) & 33.57 (0.11) \\ \hline
		\textbf{matrixssl}     & 0.61 (0.10) & 28.88 (2.54)  & 3.11 (0.09)  & 25.13 (1.33)  & 1.24 (0.32)  & 22.62 (4.16) \\ \hline
		\textbf{mbedtls}       & 0.12 (0.02) & 4.15 (0.05)   & 1.33 (0.01)  & 4.63 (0.24)   & 0.37 (0.19)  & 4.22 (0.01)  \\ \hline
		\textbf{openssl1}      & 0.59 (0.16) & 13.97 (0.72)  & 2.30 (0.82)  & 15.37 (1.48)  & 2.15 (0.92)  & 13.69 (1.18) \\ \hline
		\textbf{openssl3}      & 1.18 (0.63) & 17.94 (4.26)  & 1.77 (0.48)  & 13.14 (2.84)  & 2.32 (0.81)  & 13.67 (1.34) \\ \hline
		\textbf{openswan}      & 2.61 (0.86) & 14.06 (0.86)  & 4.55 (0.34)  & 13.80 (0.83)  & 5.39 (1.31)  & 15.67 (0.65) \\ \hline
		\textbf{strongswan}    & 0.19 (0.08) & 9.16 (2.41)   & 1.55 (0.68)  & 4.65 (0.44)   & 2.82 (0.95)  & 11.73 (4.48) \\ \hline
		\textbf{wpasupplicant} & 2.42 (0.43) & 29.18 (4.44)  & 5.45 (0.71)  & 31.59 (2.09)  & 7.61 (1.39)  & 28.05 (4.53) \\ \hline
		\textbf{Average}       & 1.45 (1.77) & 16.75 (10.58) & 3.35 (3.04)  & 17.34 (11.58) & 3.72 (3.73)  & 16.47 (9.59) \\ \hline
\end{tabular}

	\medskip
	
	\begin{tabular}{|l|c|c|c|c|c|c|}
	\hline
	\textbf{Test Subject}  & \textbf{\aflsmart} & \textbf{\honggfuzz} & \textbf{\libfuzzer} & \textbf{\nautilus} & \textbf{\peach} & \textbf{\sgfuzz} \\ \hline
	\textbf{axtls}         & 0.58 (0.23) & 0.50 (0.47) & 1.59 (0.43)  & 18.94 (2.43)  & 17.54 (0.76) & 1.24 (0.37)  \\ \hline
	\textbf{botan}         & 0.10 (0.01) & 0.00 (0.00) & 0.26 (0.15)  & 19.27 (8.91)  & 17.91 (0.02) & 0.27 (0.17)  \\ \hline
	\textbf{crypto++}      & 0.19 (0.12) & 0.02 (0.04) & 0.30 (0.13)  & 0.09 (0.01)   & 17.92 (0.03) & 0.21 (0.39)  \\ \hline
	\textbf{gnutls}        & 3.45 (0.02) & 0.01 (0.00) & 0.90 (0.42)  & 0.18 (0.00)   & 17.85 (0.07) & 0.49 (0.24)  \\ \hline
	\textbf{hostapd}       & 3.69 (0.79) & 3.64 (0.79) & 9.27 (1.04)  & 42.58 (15.77) & 17.86 (0.06) & 5.08 (0.84)  \\ \hline
	\textbf{libtomcrypt}   & 6.88 (0.10) & 7.08 (0.42) & 16.28 (1.10) & 41.69 (15.47) & 17.87 (0.04) & 12.19 (0.64) \\ \hline
	\textbf{matrixssl}     & 0.84 (0.10) & 0.40 (0.77) & 3.15 (0.49)  & 26.66 (2.33)  & 17.90 (0.03) & 3.63 (1.70)  \\ \hline
	\textbf{mbedtls}       & 0.35 (0.05) & 0.00 (0.00) & 0.93 (0.43)  & 0.18 (0.00)   & 17.89 (0.05) & 0.51 (0.17)  \\ \hline
	\textbf{openssl1}      & 0.52 (0.03) & 0.49 (0.07) & 3.05 (0.82)  & 25.71 (4.33)  & 17.89 (0.02) & 4.09 (0.32)  \\ \hline
	\textbf{openssl3}      & 0.78 (0.31) & 0.55 (0.50) & 3.67 (0.67)  & 16.78 (6.23)  & 17.86 (0.04) & 3.77 (0.38)  \\ \hline
	\textbf{openswan}      & 1.78 (0.47) & 0.44 (0.54) & 2.67 (0.52)  & 39.05 (4.68)  & 16.99 (0.01) & 3.23 (0.28)  \\ \hline
	\textbf{strongswan}    & 0.29 (0.15) & 0.00 (0.00) & 1.02 (0.46)  & 0.13 (0.04)   & 17.88 (0.02) & 0.42 (0.20)  \\ \hline
	\textbf{wpasupplicant} & 3.62 (0.80) & 3.87 (0.50) & 9.23 (1.21)  & 53.70 (10.04) & 17.89 (0.02) & 5.77 (0.53)  \\ \hline
	\textbf{Average}       & 1.78 (1.97) & 1.31 (2.10) & 4.03 (4.56)  & 21.92 (17.75) & 17.79 (0.25) & 3.15 (3.23)  \\ \hline
	\end{tabular}
\end{table*} 
\section{RQ2 Detailed Results}\label{appendix:rq2}
Tables \ref{tab:editdist} and \ref{tab:lcs} show the edit distance and normalized longest common subsequence between fuzzers' generated inputs for each test subject. Values in parentheses represent the metrics considering only the generated valid inputs.

\begin{table*}[!htbp]

	\caption{Table presenting the mean edit distance (\editdist) of inputs generated by 
	individual fuzzers across various test subjects. The reported values represent 
	the averages from five distinct fuzzing campaigns for each $\langle$fuzzer, test-subject$\rangle$ pair.
	A higher edit distance suggests a greater diversity of inputs generated by the fuzzer. 
The values in parentheses denote the mean edit distance, considering only the generated valid inputs.
	*, \eg, \afl-*, refers to fuzzing campaign with deterministic mode enabled.
	}
	\centering
	\vspace*{0.85em}
	\label{tab:editdist}
	\scriptsize
	\begin{tabular}{|l|l|l|l|l|}
		\hline
		\textbf{Test Subject}  & \textbf{\afl}    & \textbf{\afl*}  & \textbf{\aflplusplus} & \textbf{\aflplusplus*} \\ \hline
		\textbf{axtls}         & 2968.09 (57.09)  & 52.47 (16.12)   & 1682.80 (90.87)       & 127.98 (18.17)         \\ \hline
		\textbf{botan}         & 1892.97 (45.24)  & 1631.52 (12.32) & 1048.43 (62.38)       & 474.08 (19.35)         \\ \hline
		\textbf{crypto++}      & 2072.93 (76.35)  & 72.58 (11.63)   & 738.79 (91.74)        & 169.62 (11.76)         \\ \hline
		\textbf{gnutls}        & 1660.35 (37.14)  & 1462.82 (13.11) & 981.05 (43.80)        & 1162.70 (21.80)        \\ \hline
		\textbf{hostapd}       & 2964.63 (126.71) & 13.47 (11.79)   & 590.19 (128.67)       & 100.49 (15.85)         \\ \hline
		\textbf{libtomcrypt}   & 1910.82 (116.48) & 12.55 (13.69)   & 726.86 (161.45)       & 27.33 (25.20)          \\ \hline
		\textbf{matrixssl}     & 1873.86 (79.85)  & 94.62 (14.26)   & 909.81 (86.02)        & 137.23 (47.27)         \\ \hline
		\textbf{mbedtls}       & 1801.96 (50.63)  & 344.80 (10.36)  & 1257.18 (59.49)       & 498.57 (16.32)         \\ \hline
		\textbf{openssl1}      & 1738.51 (58.27)  & 111.87 (15.33)  & 1394.91 (104.51)      & 337.41 (51.37)         \\ \hline
		\textbf{openssl3}      & 1899.37 (89.63)  & 58.07 (30.50)   & 965.50 (116.94)       & 392.46 (35.78)         \\ \hline
		\textbf{openswan}      & 1952.32 (52.31)  & 1211.98 (18.47) & 1211.56 (180.77)      & 436.42 (48.48)         \\ \hline
		\textbf{strongswan}    & 1784.58 (68.98)  & 12.95 (12.15)   & 1295.49 (67.56)       & 848.10 (13.95)         \\ \hline
		\textbf{wpasupplicant} & 2428.22 (117.56) & 12.12 (11.62)   & 499.01 (113.31)       & 88.88 (18.67)          \\ \hline
		\textbf{Average}       & 2072.97 (75.09)  & 391.68 (14.72)  & 1023.20 (100.58)      & 369.33 (26.46)         \\ \hline
	\end{tabular}
	\medskip
	
	\begin{tabular}{|l|l|l|l|l|l|}
		\hline
		\textbf{Test Subject} & \textbf{\darwin} & \textbf{\darwin*} & \textbf{\aflsmart} & \textbf{\honggfuzz} & \textbf{\libfuzzer} \\ \hline
		\textbf{axtls}         & 5467.48 (150.59) & 217.76 (15.91)   & 3145.00 (53.82)    & 498.17 (169.75)     & 305.75 (111.94)     \\ \hline
		\textbf{botan}         & 4400.43 (137.00) & 1242.18 (22.85)  & 1927.80 (35.50)    & 595.19 (169.64)     & 226.42 (104.54)     \\ \hline
		\textbf{crypto++}      & 5978.43 (163.28) & 81.08 (12.68)    & 1951.84 (64.46)    & 547.32 (155.37)     & 264.98 (123.09)     \\ \hline
		\textbf{gnutls}        & 4761.86 (77.93)  & 3145.62 (21.54)  & 1835.38 (40.47)    & 202.69 (188.96)     & 189.69 (44.87)      \\ \hline
		\textbf{hostapd}       & 4751.85 (176.59) & 12.10 (11.84)    & 2309.85 (123.61)   & 688.68 (338.50)     & 293.47 (177.43)     \\ \hline
		\textbf{libtomcrypt}   & 3391.27 (182.65) & 11.97 (13.69)    & 1933.29 (115.51)   & 524.34 (344.30)     & 263.16 (168.20)     \\ \hline
		\textbf{matrixssl}     & 4692.08 (142.45) & 1291.58 (21.11)  & 1812.02 (78.88)    & 783.34 (190.01)     & 283.69 (134.80)     \\ \hline
		\textbf{mbedtls}       & 2758.39 (81.20)  & 655.77 (10.66)   & 1841.31 (49.29)    & 324.46 (161.40)     & 182.81 (70.83)      \\ \hline
		\textbf{openssl1}      & 3545.02 (130.55) & 110.44 (18.58)   & 1781.11 (57.38)    & 1221.14 (207.27)    & 294.23 (127.72)     \\ \hline
		\textbf{openssl3}      & 3921.19 (140.51) & 743.93 (21.72)   & 2393.88 (66.92)    & 709.11 (204.70)     & 299.05 (131.29)     \\ \hline
		\textbf{openswan}      & 4889.09 (90.39)  & 1705.17 (33.25)  & 1774.71 (39.18)    & 484.01 (167.22)     & 309.49 (127.78)     \\ \hline
		\textbf{strongswan}    & 3154.10 (164.76) & 16.64 (12.77)    & 1505.77 (64.83)    & 492.82 (182.52)     & 226.94 (48.66)      \\ \hline
		\textbf{wpasupplicant} & 4928.18 (192.30) & 16.94 (11.92)    & 2238.28 (109.65)   & 689.30 (346.49)     & 296.27 (177.68)     \\ \hline
		\textbf{Average}       & 4356.88 (140.78) & 711.63 (17.58)   & 2034.63 (69.19)    & 596.97 (217.39)     & 264.30 (119.14)     \\ \hline
	\end{tabular}
	\medskip

	\begin{tabular}{|l|l|l|l|l|}
		\hline
		\textbf{Test Subject}   & \textbf{\nautilus} & \textbf{\peach} & \textbf{\sgfuzz} & \textbf{\morpheus} \\ \hline
		\textbf{axtls}          & 497.38 (84.92)     & 24.93 (7.46)    & 317.14 (124.57)  & 46.76 (9.68)       \\ \hline
		\textbf{botan}          & 500.07 (6.75)      & 21.01 (7.63)    & 232.80 (62.28)   & 52.11 (9.51)       \\ \hline
		\textbf{crypto++}       & 477.94 (339.37)    & 22.88 (7.55)    & 278.29 (187.89)  & 49.57 (9.79)       \\ \hline
		\textbf{gnutls}         & 632.30 (333.35)    & 23.53 (7.51)    & 180.18 (40.20)   & 48.57 (9.43)       \\ \hline
		\textbf{hostapd}        & 430.59 (72.58)     & 21.08 (7.68)    & 281.06 (184.89)  & 52.95 (9.44)       \\ \hline
		\textbf{libtomcrypt}    & 378.91 (8.69)      & 22.05 (7.48)    & 275.52 (132.72)  & 48.14 (9.85)       \\ \hline
		\textbf{matrixssl}      & 486.73 (76.13)     & 21.41 (7.88)    & 207.46 (107.08)  & 51.17 (10.03)      \\ \hline
		\textbf{mbedtls}        & 634.60 (342.93)    & 21.02 (7.55)    & 179.42 (70.03)   & 48.05 (9.72)       \\ \hline
		\textbf{openssl1}       & 481.84 (70.25)     & 22.00 (7.48)    & 282.90 (100.41)  & 50.55 (9.72)       \\ \hline
		\textbf{openssl3}       & 521.70 (39.76)     & 21.95 (7.34)    & 288.80 (130.50)  & 52.40 (9.96)       \\ \hline
		\textbf{openswan}       & 348.61 (12.86)     & 24.14 (6.89)    & 311.27 (106.50)  & 47.13 (9.81)       \\ \hline
		\textbf{strongswan}     & 508.86 (321.44)    & 22.53 (7.44)    & 187.33 (68.14)   & 50.78 (10.06)      \\ \hline
		\textbf{wpasupplicant}  & 400.83 (83.58)     & 19.34 (7.71)    & 283.63 (179.91)  & 48.25 (9.83)       \\ \hline
		\textbf{Average}        & 484.64 (137.89)    & 22.14 (7.51)    & 254.29 (115.01)  & 49.73 (9.76)       \\ \hline
	\end{tabular}
\end{table*}
 \begin{table*}\scriptsize
	\centering
	\caption{Table presenting the mean longest common subsequence (\lcs) of inputs generated by 
	individual fuzzers across various test subjects. The reported values represent 
	the averages from five distinct fuzzing campaigns for each $\langle$fuzzer, test-subject$\rangle$ pair. 
	A lower \lcs suggests a greater diversity of inputs generated by the fuzzer. The values in parentheses denote the mean longest common subsequence, considering only the generated valid inputs.
	*, \eg, \afl-*, refers to fuzzing campaign with deterministic mode enabled.
	}
	\vspace*{0.85em}
	\label{tab:lcs}
	\begin{tabular}{|l|l|l|l|l|l|l|l|}
		\hline
		\textbf{Test Subject}  & \textbf{\afl} & \textbf{\afl*} & \textbf{\aflplusplus} & \textbf{\aflplusplus*} & \textbf{\darwin} & \textbf{\darwin*} & \textbf{\aflsmart}\\ \hline
		\textbf{axtls}         & 0.36 (0.91)   & 0.93 (0.98)    & 0.45 (0.86)           & 0.83 (0.97)            & 0.46 (0.76)      & 0.92 (0.98)       & 0.36 (0.91)       \\ \hline
		\textbf{botan}         & 0.35 (0.93)   & 0.40 (0.98)    & 0.46 (0.90)           & 0.50 (0.97)            & 0.45 (0.78)      & 0.63 (0.96)       & 0.34 (0.94)       \\ \hline
		\textbf{crypto++}      & 0.35 (0.88)   & 0.89 (0.98)    & 0.46 (0.86)           & 0.77 (0.98)            & 0.46 (0.75)      & 0.91 (0.98)       & 0.35 (0.90)       \\ \hline
		\textbf{gnutls}        & 0.31 (0.94)   & 0.34 (0.98)    & 0.35 (0.94)           & 0.42 (0.97)            & 0.42 (0.87)      & 0.43 (0.96)       & 0.44 (0.94)       \\ \hline
		\textbf{hostapd}       & 0.38 (0.81)   & 0.98 (0.98)    & 0.55 (0.81)           & 0.88 (0.98)            & 0.45 (0.73)      & 0.98 (0.98)       & 0.40 (0.80)       \\ \hline
		\textbf{libtomcrypt}   & 0.45 (0.82)   & 0.98 (0.98)    & 0.62 (0.75)           & 0.96 (0.96)            & 0.58 (0.72)      & 0.98 (0.98)       & 0.45 (0.82)       \\ \hline
		\textbf{matrixssl}     & 0.33 (0.87)   & 0.86 (0.98)    & 0.50 (0.86)           & 0.81 (0.92)            & 0.40 (0.77)      & 0.81 (0.97)       & 0.35 (0.87)       \\ \hline
		\textbf{mbedtls}       & 0.34 (0.92)   & 0.56 (0.98)    & 0.57 (0.91)           & 0.79 (0.98)            & 0.43 (0.87)      & 0.57 (0.98)       & 0.35 (0.92)       \\ \hline
		\textbf{openssl1}      & 0.35 (0.91)   & 0.91 (0.98)    & 0.45 (0.84)           & 0.64 (0.91)            & 0.48 (0.79)      & 0.88 (0.97)       & 0.35 (0.91)       \\ \hline
		\textbf{openssl3}      & 0.37 (0.86)   & 0.93 (0.95)    & 0.40 (0.82)           & 0.56 (0.94)            & 0.48 (0.77)      & 0.84 (0.97)       & 0.36 (0.89)       \\ \hline
		\textbf{openswan}      & 0.44 (0.92)   & 0.58 (0.97)    & 0.47 (0.69)           & 0.64 (0.92)            & 0.62 (0.86)      & 0.73 (0.95)       & 0.44 (0.94)       \\ \hline
		\textbf{strongswan}    & 0.36 (0.89)   & 0.98 (0.98)    & 0.41 (0.89)           & 0.49 (0.98)            & 0.47 (0.74)      & 0.98 (0.98)       & 0.37 (0.90)       \\ \hline
		\textbf{wpasupplicant} & 0.38 (0.82)   & 0.98 (0.98)    & 0.57 (0.83)           & 0.89 (0.97)            & 0.48 (0.71)      & 0.98 (0.98)       & 0.38 (0.83)       \\ \hline
		\textbf{Average}       & 0.37 (0.88)   & 0.80 (0.98)    & 0.48 (0.84)           & 0.71 (0.96)            & 0.48 (0.78)      & 0.82 (0.97)       & 0.38 (0.89)       \\ \hline
	\end{tabular}

	\medskip

	\begin{tabular}{|l|l|l|l|l|l|l|}
		\hline
		\textbf{Test Subject}   & \textbf{\honggfuzz} & \textbf{\libfuzzer} & \textbf{\nautilus} & \textbf{\peach} & \textbf{\sgfuzz} & \textbf{\morpheus} \\ \hline
		\textbf{axtls}          & 0.39 (0.72)         & 0.43 (0.82)         & 0.52 (0.85)        & 0.97 (0.99)     & 0.42 (0.81)      & 0.94 (0.98)        \\ \hline
		\textbf{botan}          & 0.39 (0.71)         & 0.40 (0.83)         & 0.53 (0.99)        & 0.98 (0.99)     & 0.41 (0.90)      & 0.93 (0.98)        \\ \hline
		\textbf{crypto++}       & 0.37 (0.75)         & 0.37 (0.80)         & 0.54 (0.45)        & 0.97 (0.99)     & 0.45 (0.72)      & 0.93 (0.98)        \\ \hline
		\textbf{gnutls}         & 0.39 (0.71)         & 0.38 (0.93)         & 0.45 (0.45)        & 0.98 (0.99)     & 0.35 (0.94)      & 0.94 (0.98)        \\ \hline
		\textbf{hostapd}        & 0.39 (0.47)         & 0.45 (0.75)         & 0.63 (0.88)        & 0.98 (0.99)     & 0.44 (0.71)      & 0.93 (0.98)        \\ \hline
		\textbf{libtomcrypt}    & 0.44 (0.46)         & 0.61 (0.75)         & 0.71 (0.99)        & 0.98 (0.99)     & 0.57 (0.80)      & 0.94 (0.98)        \\ \hline
		\textbf{matrixssl}      & 0.43 (0.70)         & 0.40 (0.78)         & 0.58 (0.87)        & 0.98 (0.99)     & 0.68 (0.84)      & 0.93 (0.98)        \\ \hline
		\textbf{mbedtls}        & 0.38 (0.74)         & 0.39 (0.89)         & 0.45 (0.44)        & 0.98 (0.99)     & 0.35 (0.89)      & 0.94 (0.98)        \\ \hline
		\textbf{openssl1}       & 0.44 (0.67)         & 0.43 (0.81)         & 0.62 (0.88)        & 0.98 (0.99)     & 0.48 (0.84)      & 0.93 (0.98)        \\ \hline
		\textbf{openssl3}       & 0.44 (0.66)         & 0.42 (0.80)         & 0.57 (0.93)        & 0.98 (0.99)     & 0.46 (0.81)      & 0.93 (0.98)        \\ \hline
		\textbf{openswan}       & 0.52 (0.73)         & 0.45 (0.79)         & 0.69 (0.98)        & 0.98 (0.99)     & 0.42 (0.83)      & 0.94 (0.98)        \\ \hline
		\textbf{strongswan}     & 0.39 (0.70)         & 0.37 (0.92)         & 0.57 (0.48)        & 0.98 (0.99)     & 0.37 (0.90)      & 0.93 (0.98)        \\ \hline
		\textbf{wpasupplicant}  & 0.38 (0.45)         & 0.45 (0.74)         & 0.67 (0.87)        & 0.98 (0.99)     & 0.45 (0.73)      & 0.94 (0.98)        \\ \hline
		\textbf{Average}        & 0.41 (0.65)         & 0.43 (0.82)         & 0.58 (0.77)        & 0.98 (0.99)     & 0.45 (0.82)      & 0.94 (0.98)        \\ \hline
	\end{tabular}
\end{table*}

\end{document}